# Comparison of mechanical conditions in a lower leg model with 5 or 6 tissue types while exposed to prosthetic sockets applying finite element analysis

Sara Kallin[a,b], Asim Rashid[a], Kent Salomonsson[a], Peter Hansbo[a]

[a]Jönköping University, Dept of Product Development, School of Engineering, Jönköping, Sweden. [b] Jönköping University, Dept of Rehabilitation, School of Health and Welfare, Jönköping, Sweden.
Corresponding author: Sara Kallin, sara.kallin@ju.se , address: Jönköping University, Box 1026, S-551 11 Jönköping, Sweden.

## Abstract

Lower limb amputees often suffer skin and tissue problems from using their prosthesis which is a challenging biomechanical problem. The finite element method (FEM) has previously been applied to analyse internal mechanical conditions of the leg at prosthesis use. However, the representation of soft tissue was simplified to few layers and tissue types. The effects of such a simplification of human tissue is still unclear and the results from simplified models may be misleading. Thus, comparisons of the effects of using five versus six tissue types were performed on a transtibial cross section model exposed to three different socket designs. Skin, fat, vessels and bones were defined separately while muscle and fascia tissues were separate or merged. Nonlinear behaviour and friction between socket and skin were considered in the simulations. Contact forces as well as internal stresses and strains of each tissue type differed in both magnitude and maxima site for each material set within and between the different prosthetic socket conditions. Relative changes of stresses and strains by several hundred percent were found when fascia and muscle material properties were merged compared to when they were modelled separately. Thus, the level of tissue detail needs to be considered when creating limb models and interpreting results of related FEM simulations.

# 1 Introduction

Lower limb amputees who use a prosthesis often suffer blisters, oedema, skin irritation, dermatitis and pressure ulcers [1, 2]. The load on the residual limb is unavoidable and biomechanical understanding of the interaction between the human tissue and the prosthetic socket is important in order to design a comfortable and practical socket with proper load distribution. [1]. Fergason &Smith defined three concepts of socket design used in a transtibial prosthesis, based on previous developments: total contact (TC), total surface-bearing (TSB) and hydrostatic (HS) sockets [3]. Liners, socks and padding may be used between the skin and the hard plastic socket depending on the concepts used, and individual adjustment. Despite these efforts of distributing loads on the residual limb, 15-82 % of lower limb amputees experience skin problems [4-6]. A person using a prosthetic socket is also at a risk of developing deep tissue injury (DTI) [7, 8]. Excessive levels of tissue strains and stresses have been found at bony prominences [8], and support the theory that DTI starts in the deep tissues underneath the intact skin [9, 10]. Thus, it is important to consider the internal mechanical conditions in the soft tissues while designing a prosthetic socket.

Finite element analysis (FEA) has been used in many studies to simulate the behaviour of soft tissues under external loading [7, 11-13]. Given realistic boundary conditions, geometries and material models, FEA can be used to predict the stress-strain distribution and enable parametric studies to optimize the design of sockets. In previous works, finite element (FE) models were employed to investigate contact stresses at the skin-socket interface in an attempt to optimize the design of a prosthetic socket [1, 14]. The internal mechanical conditions in the limb were ignored. In more recent works, internal mechanical conditions in the lower limb under external loading were studied [7, 8, 15-18]. The different layers of soft tissues were merged (lumped) together, most commonly into three tissue types. When merging tissue layers for representation with the same properties typically the tissue types are assigned with same properties from one of the types, e.g. merging/lumping geometries of muscles, connective tissue, vessels and nerves assigning all with muscle properties. The development of models with more detailed tissue geometries in the human lower limb have recently been published, e.g., cross section of tibial compartments with explicit skin, fat, fascia and muscles [19], and individual hamstrings, quadriceps and gluteal muscles [20]. Moerman *et al.* investigated the influence of individual gluteal geometries on tissue loads with FEA, using one set of material properties for the soft tissue segments skin, adipose fat and muscles [21]. They showed that damage risk volume (DRV) - volume of tissues exposed to shear strains above 50% - were geometry dependent and recommended future studies to further investigate the relative importance of varying material compositions and/or geometries. Material behaviour in simulated human soft tissue has been represented by different material models and related material parameters, e.g., linear elastic models [22, 23], nonlinear by polynomial models [24, 25] and hyperelastic by elastic strain energy functions such as James-Green-Simpson [26, 27], Neo-Hookean [8, 28-32] and Ogden models [16, 20, 33]. The diverse use of models reflects the complexity of human soft tissue behaviour and the need for further development. Comparison of material models on soft tissue level is rare but is important in evaluating FE simulations due to soft tissue complexity and model limitations [13]. There is also a lack of comparison between different material representations for the same loading situation.

The aim of this study is to examine the mechanical consequences of combining fascia and muscle materials for respective regions compared to representing them separately, in a generic transverse cross section model of a lower limb exposed to three different prosthetic socket designs.

Specifically, we aim to investigate mechanical conditions by internal stresses and strains, and by contact pressure between skin and socket models, simulating a section at the middle height of the calf in a non-weightbearing lower limb contained in different prosthetic sockets. A detailed FE model of a generic transtibial transversal section will be developed. Skin, fat, fascia, muscles, blood vessels and bones will be represented separately and material properties obtained from the literature. The contact interaction between the skin and socket will include friction. The resulting stresses, strains and contact pressures using two sets of material properties will be compared under the three socket design conditions..

## 2 Method

### 2.1 Model

The cross section of the lower limb considered in this study (Figure 1) is based on an anatomy illustration of the horizontal cross section just above the midpoint of the calf-height [34], since geometries of the tissues are predetermined and easily segmented. The geometries of the anatomical features skin, fat, fasciae, separate muscles, vessels and bones were set up accordingly. Nerves and lymphatic vessels were ignored. The geometries of three sockets: Total contact (TC), Total surface bearing (TSB) and Hydrostatic (HS) used in the simulation were defined according to the literature [3] and clinical practice. For the HS socket, the encapsulated limb area was reduced by 5% according to clinical practice of a 3-6% volume reduction. Three models were created, one for each limb-socket concept, with 4.5mm socket thickness. The overlay of limb and socket geometries is displayed in Figure 2.

The assembly of the limb contained 26 330 elements and each socket cross section contained around 3000 elements. The elements used for the parts' meshes were 8-node biquadratic plane strain quadrilateral, hybrid, linear pressure elements, and 6-node quadratic plane strain triangle, hybrid, linear pressure elements, respectively (CPE8H and CPE6H in Abaqus 6.14-3, Dassault Systèmes).

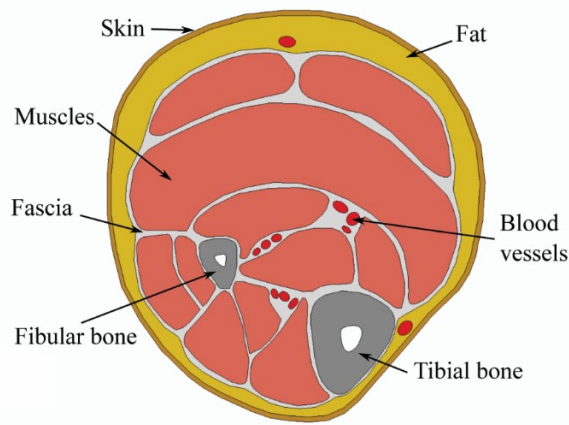

**Figure 1** Lower limb cross section model of the calf based on Netter (1987) [34] (page 98).

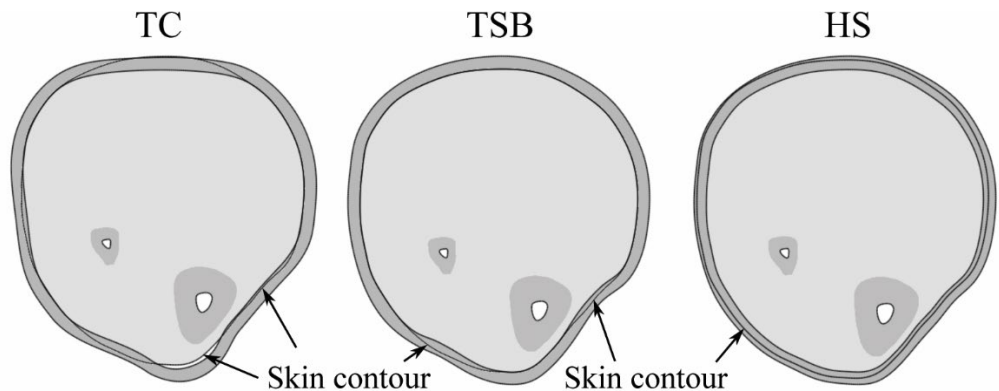

**Figure 2** The cross section geometries of the limb and socket models overlap. Socket concepts: total contact (TC), total surface bearing (TSB), and hydrostatic (HS) sockets. The limb is represented by light grey, bones and sockets by dark grey.

## 2.2 Material properties

Two material sets were considered; one with separate material properties of each tissue type (Separate set) and one set where muscle and fasciae regions had the same combined material properties while the other tissues remained separate (Combined set). For the second material set used in the current study, (Combined set), the properties of fascia and muscle were combined and calibrated to a new combined material property (see section 2.2.1 for this procedure) while other tissues remained unchanged from the Separate material set.

Soft tissues were modelled as hyperelastic isotropic materials by strain energy potentials in Ogden- or Yeoh-form taking nonlinear behaviour and large deformation into consideration (eq. (1) – (2)). The Ogden model for slightly compressible materials is considered good for large deformation and used to represent the behaviour of human tissues. The material data used was retrieved from literature (Table 1). We chose material parameters based primarily on experimental data from compression tests and in transverse plane if possible, to comply with the chosen 2D plane. Data were chosen from studies applying the Ogden model where such were available to keep variations between material models low. Skin, fat and muscle tissues were represented by the Ogden first order model, and fasciae by the Yeoh form. Bone and blood vessels were modelled as linearly elastic materials.

Skin mechanical properties vary between living and dead human tissue, by body site [35] and between humans and animals [36]. For the purpose of this study, the skin was assumed isotropic and nearly incompressible, modelled as a first order Ogden hyperelastic material retrieved from Payne *et al* [37] based on compressive strain data from Shergold *et al* [38], and with $D_1$ calculated from $\mu_1$ and Poisson's ratio of 0.495 [39]. Nonlinear behaviour of subcutaneous fat, the adipose tissue, was modelled by adapting the first order Ogden material model according to Payne *et al* [37] based on compressive strain from Comley & Fleck [40] while assuming isotropic and hyperelastic properties. $D_1$ was calculated as described above. In the present study, muscle tissue was assumed to be isotropic and hyperelastic and modelled as a first order Ogden material with slight compressibility, and parameters from Al Dirini *et al* [20]. Fasciae material data were retrieved by extracting data points from Figure 3 in Pavan *et al* [19] which was based on human deceased lower limb anterior and posterior fasciae in the mediolateral direction. The mean of anterior and posterior fasciae data was then used to determine material parameters and coefficients by a curve fitting process performed using the free software Hyperfit 2.X (Faculty of Mechanical Engineering, Institute of solid Mechanics, Mechatronics and Biomechanics, Brno University of Technology, Czech Republic).

Vessels and blood were represented together as a linear elastic material with a modulus of elasticity of 10 kPa and Poisson's ratio of 0.49. These properties depend on blood pressure and volume. With a modulus of elasticity of 10 kPa, a maximum pressure of 14.7 kPa was predicted in the vessels after contact was established between socket and skin. This pressure, equivalent to 110 mmHg, was in the range of normal blood pressure. The tibia bone was assumed isotropic with a Young's modulus of 11.8 GPa from Hoffmeister *et al.* [41] based on transverse loading of the tibia. The fibula bone was assumed to have the same properties. The prosthetic socket was assumed to be made of a material used for prosthetic temporary test-sockets: glycol modified polyethylene terephthalate, PETG, with a modulus of elasticity of 459 MPa [42] and Poisson's ratio of 0.4.

**Table 1 Material properties used in the models**

| Material for tissue region | Material model | Parameters and coefficients | Reference |
|---|---|---|---|
| Skin | Ogden 1st (Equation (1)) | $\mu_1 = 220.0\ kPa$, $\alpha_1 = 12$ $D_1 = 9.12 * 10^{-8}\ Pa^{-1}$ | [37], [38] |
| Fat | Ogden 1st (Equation (1)) | $\mu_1 = 1.700\ kPa$, $\alpha_1 = 26$ $D_1 = 1.18 * 10^{-5}\ Pa^{-1}$ | [37], [40] |
| Muscle* | Ogden 1st (Equation (1)) | $\mu_1 = 1.907\ kPa$, $\alpha_1 = 4.6$ $D_1 = 1.05 * 10^{-5}\ Pa^{-1}$ | [20] |
| Fascia* | Yeoh (Equation (2)) | $C_{10} = 4.91\ MPa$ $C_{20} = 13.59\ MPa$ $C_{30} = 18.97\ MPa$ | Modified from [19] |
| Combined Muscle-Fascia** | Ogden 1st (Equation (1)) | $\mu_1 = 12.0\ kPa$, $\alpha_1 = 14$ $D_1 = 1.67 * 10^{-6}\ Pa^{-1}$ | |
| Blood vessels | Linear, elastic | $E = 10.00\ kPa$ $\nu = 0.49$ | |
| Cortical bone | Linear, elastic | $E = 11.80\ GPa$ $\nu = 0.394$ | [41] |
| Sockets, PETG | Linear, elastic | $E = 459\ MPa$ $\nu = 0.4$ | [42] |

*Note: $D_1$ is computed from $\mu_1$ and $\nu=0.495$, according to equation 3.273, p.85, chapter 3 in Silber & Then [39].*
*\*Separate material properties used in the Separate set, \*\*combined material properties used for the fascia and muscle regions in the Combined set.*

The Ogden material model was chosen for the combined material since muscle tissue represented the largest area in the model.

2.2.1 Combined material calibration

The parameters for the combined material were determined by a designed indentation test simulation for calibration of the compound behaviour of the two material sets by fitting of their reaction force– displacement curves (Figure 3). The coefficients ($\mu$, $\alpha$ and $D$) for the combined material of fascia and muscle were determined by a heuristic optimization process, systematically altering them while stepwise evaluating the reaction force-displacement curves to the reference curve (right graph in Figure 3). The Separate material set including 6 tissues, was used as the reference and represented by the solid line in Figure 3 (right). The second material set with the final combined fascia and muscle properties is represented by the line-dot-line curve. This new material was termed Combined set and is presented in Table 1 as *Combined muscle-fasciae*. The common strategy of neglecting fascia was also tested by assigning fascia with muscle tissue properties in the material calibration for comparison. That compound response is also displayed in Figure 3, by a dotted line.

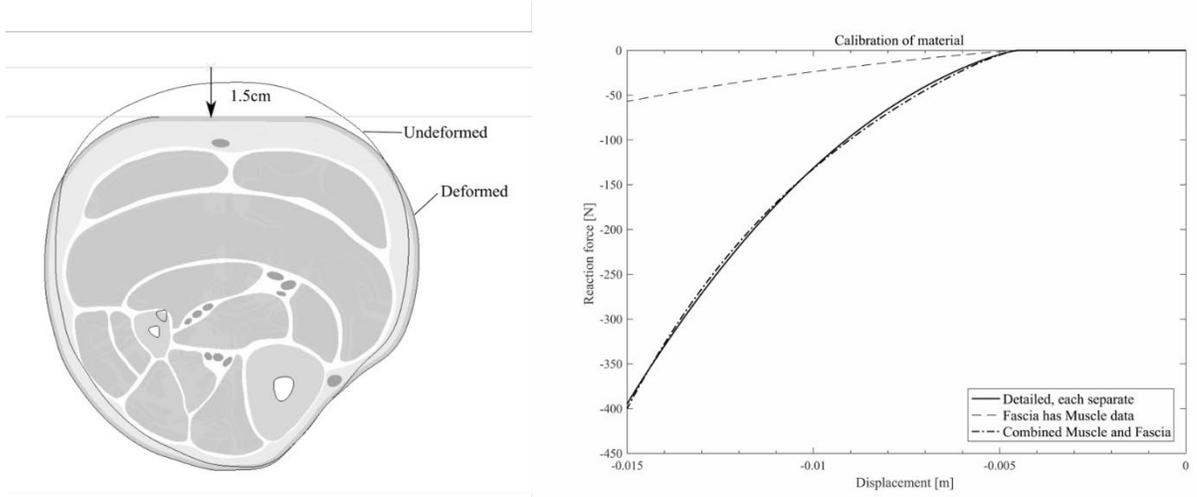

**Figure 3** Calibration of Combined Muscle-Fascia material. Left: Calibration model and applied displacement; Right: Reaction force-displacement curves with varying properties for muscle and fascia.

The Ogden strain energy potential is expressed as

$$U = \sum_{i=1}^{N} \frac{\mu_i}{\alpha_i}(\lambda_1^{\alpha_i} + \lambda_2^{\alpha_i} + \lambda_3^{\alpha_i} - 3) + \sum_{i=1}^{N} \frac{1}{D_i}(J^{el} - 1)^{2i} \quad (1)$$

where $U$ is the strain energy per unit of reference volume, $\mu_i$, $\alpha_i$ and $D_i$ are material constants, $N$ is the order of the equation, $\lambda_i$ are the deviatoric principal stretches and $J^{el}$ is the elastic volume ratio.

The polynomial strain energy potential is expressed in Yeoh form as

$$U = \sum_{i=1}^{3} C_{i0}(I_1 - 3)^i \quad (2)$$

when incompressibility is assumed. The strain energy per unit of reference volume is denoted $U$, $C_{i0}$ are material dependent parameters, $I_1$ is the first deviatoric strain invariant as $I_1 = \lambda_1^2 + \lambda_2^2 + \lambda_3^2$, $\bar{I}_1 = J^{-2/3} I_1$, where $J$ is the total volume ratio, and $\lambda_i$ are the principal stretches.

## 2.3 Boundary conditions

The socket models were assumed to be stiff test-sockets, without liner or cushioning material or limb-prosthesis-fixation. No additional loads were applied, i.e. non-weight bearing condition was simulated, however, the socket geometry constraint generated load. For all the design concepts, the initial size of the socket was smaller than the outer boundary of the limb cross-section (Figure 2), which resulted in an overlapping of the two bodies. The contact algorithm gradually resolved the overclosure over multiple increments. The contact surfaces were pushed apart until there was no more penetration. As the overclosures were resolved, stresses and strains appeared in the socket and the limb cross section. Surface-to-surface discretization was used which considered the shape of both the slave and master surfaces in the region of contact. The coefficient of friction (CoF) between the skin and the socket was assumed to be 0.4 [43]. The internal soft tissue sections shared nodes and thus no other contact conditions were applied. The inner boundary of the tibia was fixed while the fibula was free to move, which occurs in a transtibial amputation without a bone bridge [44, 45].

### 2.4 Finite element analysis and outcomes

Abaqus 6.14-3/Standard (an implicit solver) (Dassault Systèmes), was used for building the model and running analyses. For the two-dimensional simulation, the assumption of plane strain was applied.

In total, six FEA runs were performed for comparison. For each socket condition, one run with each material set (Separate or Combined) was applied to the limb model. Outcome measures chosen for comparison were contact pressure at the skin-socket interface, and effective stress (Mises), shear stress, logarithmic strains by absolute maximum principal strain (AMP), and shear strain for each tissue type. Investigations of the distribution visualisations (e.g. Figure 4 and Figure 7) for sites of maxima and minima were performed by zooming and applying limits of the outcome of interest. These investigations identified outlier/extreme values or singular nodal values to ignore, determined levels for approximate maximum and minimum magnitudes, and identified the sites of absolute maxima per tissue type and run, denoted as Max Separate and Max Combined sites. The level of approximated maximum absolute value included at least five nodes and two element neighbours. Comparisons of effects by different material sets for each socket design were performed pairwise per absolute maxima and per site, by relative change in percent (eq. 3), with the Separate material set condition results used as the reference.

$$Relative\ change(x, x_{reference}) = \frac{x - x_{reference}}{|x_{reference}|} \qquad (3)$$

# 3 Results

## 3.1 Comparison of stresses and strains

The internal mechanical stresses and strains changed in the tissues when muscle and fasciae properties were combined as compared to when separate material properties were used for each of the five soft tissue types (Table 2 and Table 4). The sites of absolute maximum magnitudes per tissue type differed between the material sets and socket designs, examples of distributions given by effective stress in Figure 4, and shear strain in Figure 7. The sites of maxima for effective stresses were found in the fascia for the Separate set and in the skin for the Combined set, regardless of socket design. The sites of maximum strains were found in muscles with the Separate set for all three socket designs. For the Combined set, maxima were found in fat for TC and HS socket designs, while maxima site for maximal principal strain was found in muscles and shear strain in vessels with the TSB socket design.

The relative change of maximum magnitudes of stresses per tissue type and socket condition increased the most by 459% (effective stress in muscles with TC socket) and decreased the most by -599% (shear stress in muscles with TC socket) in the Combined set compared to the Separate set, presented in Table 2 and Figure 5. The relative change of maximum magnitudes of strains increased the most by 202% (for shear strain in fat with HS socket) and decreased the most by -253% (for shear strain in vessels with TC and HS sockets) using the Combined set compared to the Separate set, presented in Table 3 and Figure 5.

The relative changes in stresses and strains at their maximum-sites-per-condition comparison (denoted site Max Separate and Max Combined) varied even more than the change per maximum, and are presented in Table 2, Table 3, Figure 6, and Figure 8.

## 3.2 Comparison of contact pressures

The computed skin-socket interface contact pressure changed in magnitude per material set used, and by site of maximum by socket condition. The relative change of contact pressure maximum increased with the Combined set in all socket designs, by 134%, 58% and 37% for TC, TSB and HC socket, respectively. The distribution of contact pressures at the skin-socket interface for the six conditions are displayed in Figure 9.

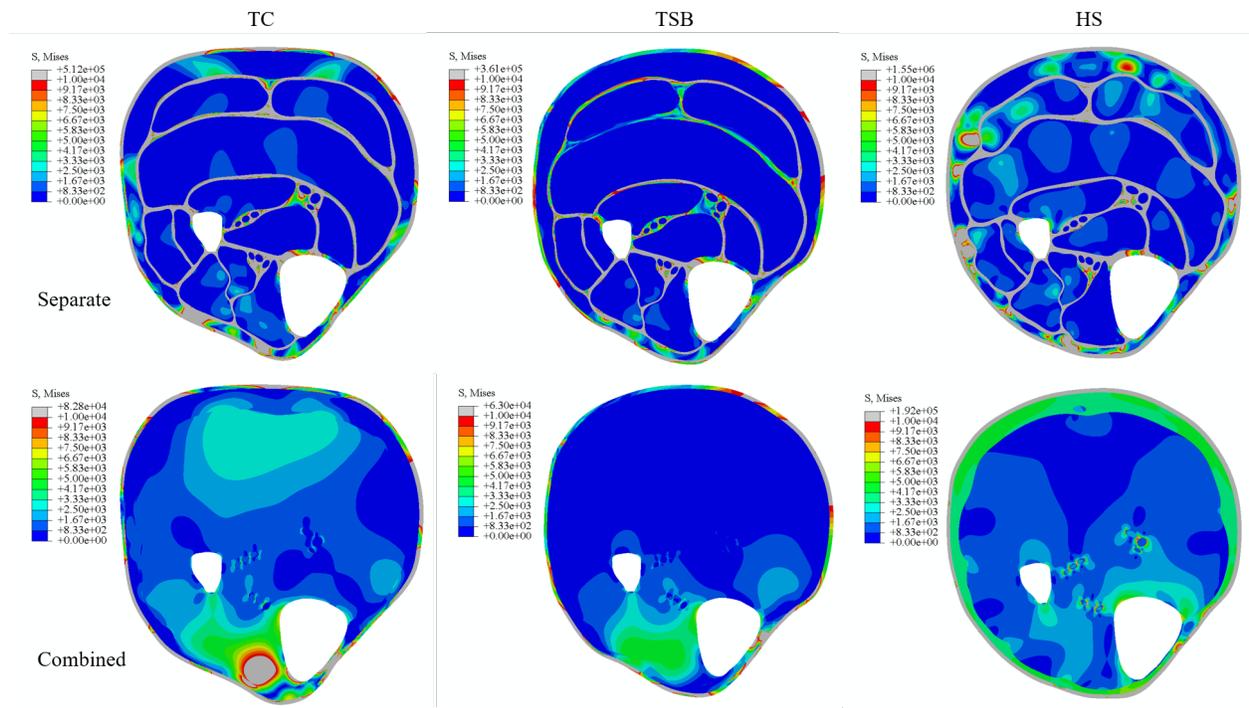

**Figure 4** Stress distribution (effective stress, Mises) in soft tissues per material set and socket design. Upper row: Separate set, lower row: Combined set. Socket designs: TC Total Contact, TSB Total Surface Bearing, HS Hydrostatic. Coloured scale limited to 0-10 kPa.

**Table 2 Stress maximum magnitudes per condition**

Stress Maxima [kPa] — Conditions: Socket type, Material set

| Tissue | | TC Separate | TC Combined | TSB Separate | TSB Combined | HS Separate | HS Combined |
|---|---|---|---|---|---|---|---|
| skin | Effective | **51.2** | 77.4 | 53.0 | 59.7 | 155.0 | **178.0** |
| | Shear abs | 28.6 | 40.1 | **-23.2** | 26.8 | **-73.0** | -68.5 |
| fat | Effective | 27.5 | 32.2 | 12.8 | 15.2 | **35.0** | 12.5 |
| | Shear abs | -10.8 | 16.16 | **-2.68** | 3.6 | **-40.0** | 6.9 |
| muscles | Effective | 2.6 | **14.53** | **1.87** | 5.1 | 5.8 | 4.5 |
| | Shear abs | -1.09 | **-7.62** | **-1.04** | -2.76 | -2.95 | -2.27 |
| fasciae | Effective | 450.0 | 32.6 | 270.0 | **18.2** | **1120** | 25.0 |
| | Shear abs | -190.0 | -6.55 | -140.0 | **-3.5** | **500.0** | -5.0 |
| vessels | Effective | **0.91** | 1.23 | 2.18 | **3.2** | 1.57 | 2.7 |
| | Shear abs | 0.398 | -0.6 | 0.69 | **1.26** | **-0.28** | -0.9 |

*Note: Range limits per outcome and tissue type marked in bold (regardless condition), based on absolute value.*
*Negative values: the stress is in opposite direction to those with positive values*

**Table 3 Stress Relative change for Combined in comparison to Separate material set**

| Tissue | Site | TC Effective stress | TC Shear stress | TSB Effective stress | TSB Shear stress | HS Effective stress | HS Shear stress |
|---|---|---|---|---|---|---|---|
| Skin | Max Separate | 51% | 40% | 13% | 216% | -5% | 6% |
| | Max Combined | 51% | 40% | 13% | 216% | 78% | 6% |
| | Max both | 51% | 40% | 13% | 216% | 15% | 6% |
| Fat | Max Separate | -73% | 74% | -7% | 34% | -89% | 100% |
| | Max Combined | 367% | 294% | 124% | 152% | 54% | 68% |
| | Max both | 17% | 250% | 19% | 234% | -64% | 117% |
| Muscles | Max Separate | 40% | -212% | 164% | -160% | -79% | 112% |
| | Max Combined | 837% | -853% | 507% | -245% | 1224% | -22800% |
| | Max both | 459% | -599% | 173% | -165% | 23% | 23% |
| Fasciae | Max Separate | -99% | 99% | -99% | 99% | -100% | -100% |
| | Max Combined | 270% | 92% | 78% | 49% | -40% | 88% |
| | Max both | -93% | 97% | -93% | 98% | -98% | -99% |
| Vessels | Max Separate | 0% | -22% | 47% | 83% | 2% | 132% |
| | Max Combined | 339% | -445% | 47% | 83% | 938% | -1400% |
| | Max both | 35% | -251% | 47% | 83% | 72% | -221% |

*Note: TC Total Contact, TSB Total Surface Bearing, HS Hydrostatic.*

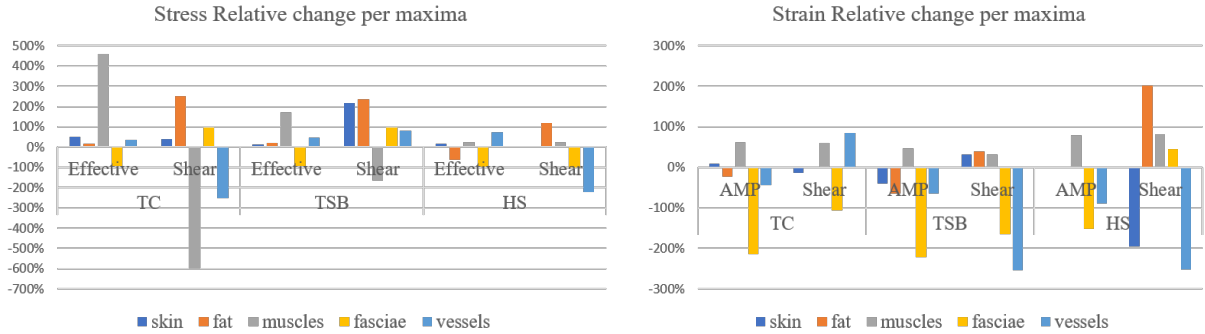

**Figure 5** Relative change, pairwise comparison of maximum magnitudes per tissue type. Relative change in percent for Combined set compared to Separate set. Relative change = (Combined – Separate)/abs(Separate) An increase with Combined set is displayed by positive change. Left: Stresses Effective stress, shear stress, Right: Strains AMP absolute value of maximal principal strain, shear strain. TC Total Contact, TSB Total Surface Bearing, HS Hydrostatic.

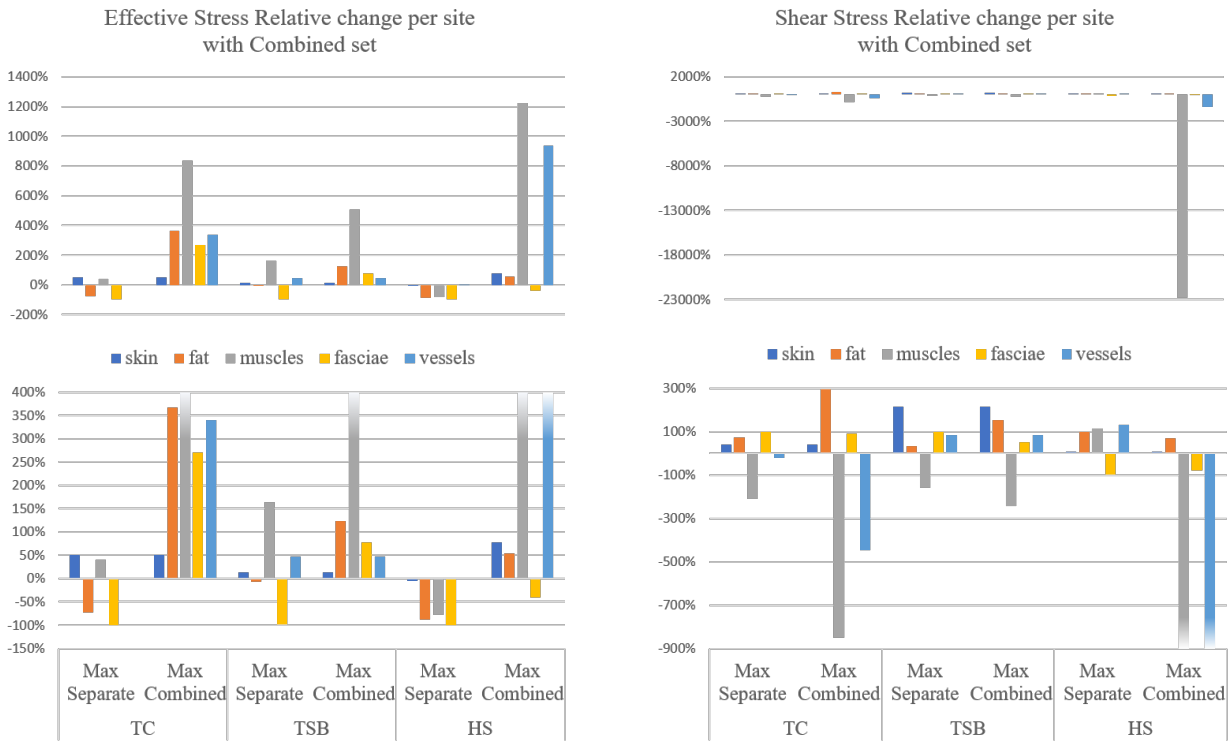

**Figure 6** Pairwise comparison of stresses per site of maximum magnitudes. Upper panels display full ranges, lower panels show limited ranges. Relative change in percent for Combined set compared to Separate set. Relative change = (Combined – Separate)/abs(Separate). An increase with Combined set is displayed by positive change. Left: Effective Stress. Right: Shear stress. TC Total Contact, TSB Total Surface Bearing, HS Hydrostatic. Max Separate: at site for maximum with the Separate set. Max Combined: at site for maximum with the Combined set.

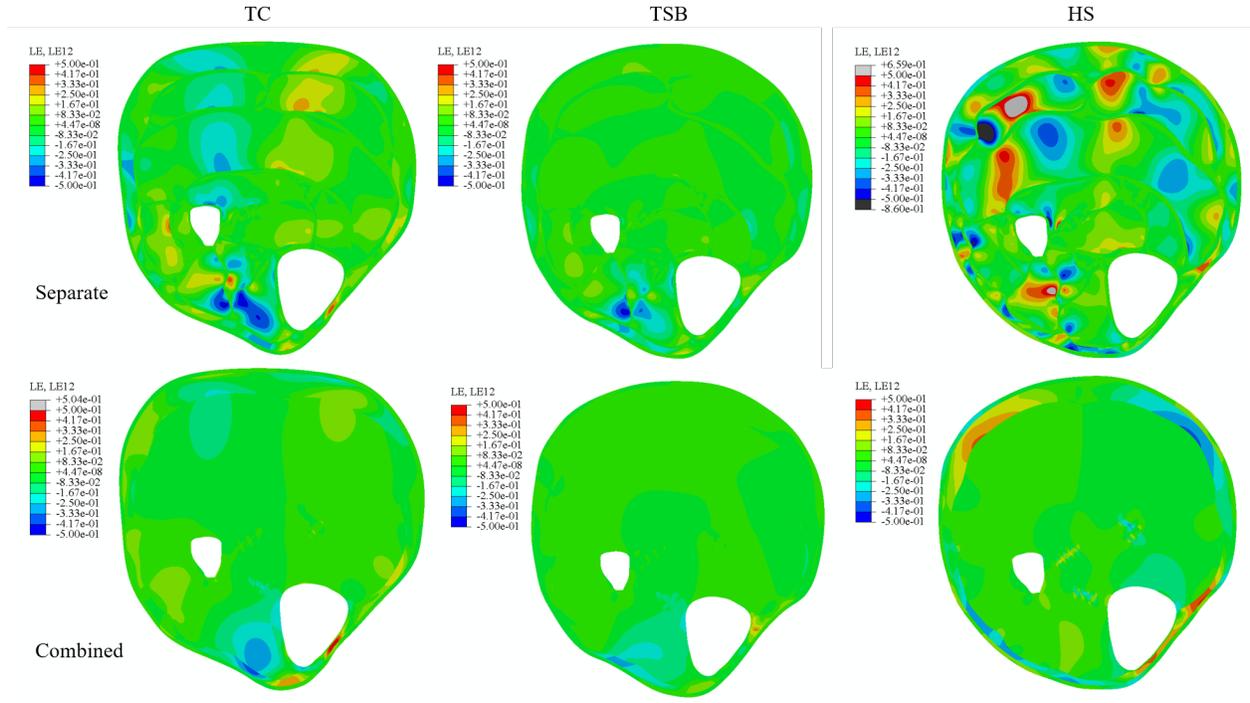

**Figure 7** Strain distribution (logarithmic shear strain) in soft tissues per material set and socket design. Upper row: Separate set, lower row: Combined set. Socket designs: TC Total Contact, TSB Total Surface Bearing, HS Hydrostatic. MPA maximum principal absolute value, LE12 shear strain. Coloured scale limited to ±50%.

**Table 4 Strain maximum per condition**

| Strain Maxima [%] | | Conditions: Socket type, Material set | | | | | |
|---|---|---|---|---|---|---|---|
| | | TC | | TSB | | HS | |
| **Tissue** | | Separate | Combined | Separate | Combined | Separate | Combined |
| skin | AMP | **-6.00** | -8.4 | 6.3 | 6.91 | **-13.4** | -13.2 |
| | Shear | 11.90 | 15.77 | **-9.6** | -10.95 | **24.1** | -23.2 |
| fat | AMP | -26.20 | **-42.97** | **-21.1** | -25.92 | -40.1 | -40.2 |
| | Shear | 35.80 | **49.85** | **-29.5** | -29.77 | -43.5 | 44.3 |
| muscles | AMP | -31.60 | -16.87 | -24.5 | **-9.44** | **-49.5** | -10.8 |
| | Shear | -46.00 | -31.72 | -45.6 | -18.04 | **-84.1** | **-16.0** |
| fasciae | AMP | -7.10 | -22.92 | **-5.7** | -17.94 | -11.5 | **-29.0** |
| | Shear | -11.10 | **-29.34** | **-10.0** | -20.74 | 18.5 | 27.0 |
| vessels | AMP | **-10.70** | -17.71 | -21.9 | -31.49 | -19.6 | **-37** |
| | Shear | 11.80 | -18.1 | 19.2 | **35.66** | **-8.0** | -28.2 |

*Note: Range limits per outcome and tissue type marked in bold (regardless condition), based on absolute value.*
*AMP: Absolute Maximum Principal. Negative values: the strain is in opposite direction to those with positive values.*

**Table 5 Strain Relative change for Combined in comparison to Separate**

| Tissue | Site | TC AMP | TC Shear | TSB AMP | TSB Shear | HS MPA | HS Shear |
|---|---|---|---|---|---|---|---|
| Skin | Max Separate | -15% | 33% | 10% | -14% | 1% | -196% |
|  | Max Combined | -250% | 33% | 10% | -14% | 1% | -196% |
|  | Max both | -40% | 33% | 10% | -14% | 1% | -196% |
| Fat | Max Separate | -55% | 37% | -17% | 12% | 23% | 78% |
|  | Max Combined | -69% | 39% | -23% | -57% | -36% | 28% |
|  | Max both | -64% | 39% | -23% | -1% | 0% | 202% |
| Muscles | Max Separate | 74% | 54% | 62% | 61% | 89% | 103% |
|  | Max Combined | 21% | 11% | 35% | 51% | -10% | -15900% |
|  | Max both | 47% | 31% | 61% | 60% | 78% | 81% |
| Fasciae | Max Separate | -10% | -47% | -60% | -29% | 23% | -45% |
|  | Max Combined | -12833% | -1360% | -9070% | -4409% | -1511% | 1250% |
|  | Max both | -223% | -164% | -215% | -107% | -152% | 46% |
| Vessels | Max Separate | -12% | -32% | -44% | 79% | -34% | 120% |
|  | Max Combined | -172% | -417% | -44% | 93% | -874% | -1310% |
|  | Max both | -66% | -253% | -44% | 86% | -89% | -253% |

*Note: AMP Absolute Maximum Principal strain. TC Total Contact, TSB Total Surface Bearing, HS Hydrostatic socket design*

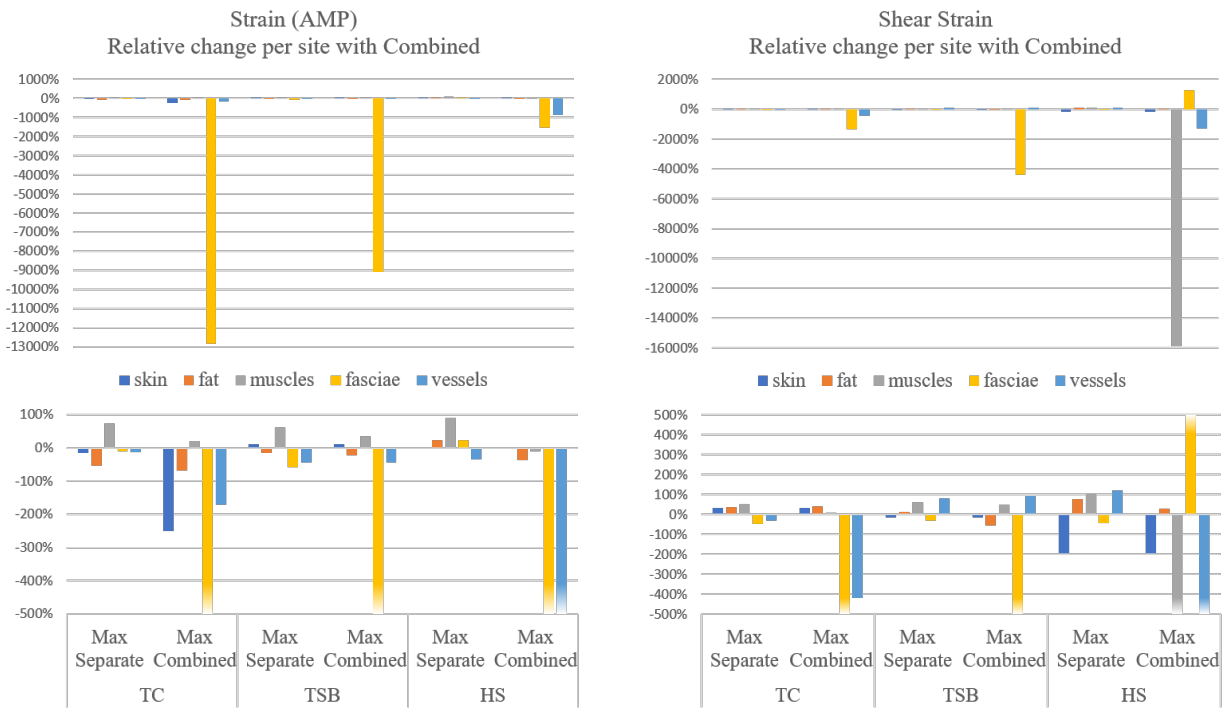

**Figure 8** Strains. Pairwise comparison of logarithmic strains per site of maximum magnitudes by relative change. in percent. Upper panels display full ranges, lower panels show limited ranges. for Combined set compared to Separate set. Relative change = (Combined – Separate)/abs(Separate). An increase with Combined set is displayed by positive change. Left: Strain Absolute Maximum Principal, AMP. Right: Shear strain. TC Total Contact, TSB Total Surface Bearing, HS Hydrostatic. Max Separate: at site for maximum with the Separate set. Max Combined: at site for maximum with the Combined set.

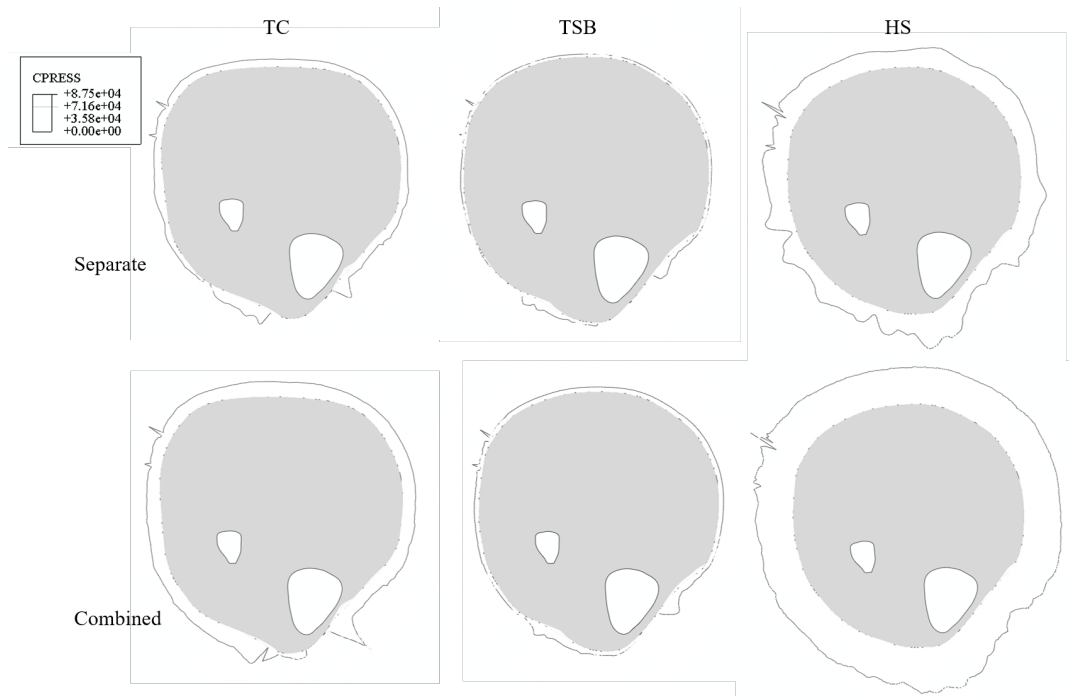

**Figure 9** Contact pressure distribution (CPRESS) [Pa] at skin-socket interface with the three socket designs and two material sets, Separate and Combined. The magnitude of the contact pressure is represented by the line and its distance from the grey limb model. TC Total contact, TSB Total surface bearing and HS Hydrostatic. (The spikes at lateral/posterior region were excluded from comparison, as they are singular nodal values.)

# 4 Discussion

This study presented FEA simulations on a lower limb model with six defined internal tissue types in 2D, and applied material data retrieved from the literature. Two material sets were applied, where the only difference was that fasciae and muscles were either given separate material properties (Separate set) or given the same material property as a combination of the two (Combined set). Loading conditions were three different prosthetic socket geometries applied without weightbearing. The resulting internal stresses, strains and skin contact pressure outcomes were compared for differences between the material sets used. The main results showed that the mechanical conditions of the tissues varied in maximum magnitude and sites of maxima by both material set used, and socket design. The differences were considerable, with relative changes often reaching several hundred percent. Both an increase and decrease in the outcomes of tissue types under the given loading conditions appeared when the Combined material set was used instead of the Separate set.

The Separate material set resulted in the largest and high stresses in the fasciae region, while the largest strains were found in the muscles. With the Combined material set, the largest stresses were found in skin and the largest strains were found in fat for the TC and HS socket designs, and in muscle and vessels for the TSB socket. In the display of shear strain distribution (Figure 7), the limits were set to ±50%, since a damage risk volume (DRV) measure for muscles at 50% strain level was presented by Moerman *et al* [21]. Shear strain absolute values found in the present study were below this threshold with the Combined set, while with the Separate set, distinct regions reached more than 50% in the HS condition, with a maximum of 84% shear strain in musculus gastrocnemius lateral head. Stecco *et al.* [46] reports a damage level of 27% strain for human leg fasciae under tensile tests, while no damage level for shear strain was presented. Our results showed maximum strain in fascia regions for the Combined set to be above this level (29% with TC and HS socket designs), while for the Separate set, maximum strain was below (largest found was with HS at 18%). Thus, a damage risk evaluation based on these simulations would differ between the two material sets used. In addition, Loerraker *et al.* emphasized the individual variation in damage thresholds due to tissue status, pathology and other factors, including ischemia [47]. Thus, there is a need to establish damage thresholds in humans for different tissue types and individual health conditions for prevention of such damage, and this should be considered in interpretations of simulations.

The current study used more detailed geometries than presented in previous lower limb models, with segments of 6 tissue types over 27 regions, whereof five types were of soft tissue. We used an anatomical illustration for the geometry, which provided clearly identifiable tissue regions and could be considered as a generic standard geometry model. Since the geometries are important [21, 48], we consider it useful to apply a standard model when comparing material property models and results and suggest this for future sensitivity studies. However, no such standard model has been proposed before.

We chose to use material properties obtained during compression and/or the selected direction of cross section, to be consistent with the simulation. The strategy used with material data from other studies has previously been applied to FEA for clinical purposes. Material properties used here, originating from different sources such as the literature, reversed FEA, etc., will lack preciseness compared to, say, careful MRI measurements and material parameter experiments on an individual basis. However, such preciseness is beyond the scope of the current paper, the purpose of which was merely to show that the increased detail in FE-geometry and property allocation will provide different information about internal conditions compared to a model with merged, lumped tissues. Interpretations of results from this and other studies should consider the modelling of the tissues.

The fascia was not ignored in either of the models, which has often been the case in previous studies. Fascia properties were obtained by curve fitting of published experimental data to different material models (see Chapter 2.2). The searched parameters for the Ogden model did not converge, while the parameters for the Yeoh model was determined quickly. Hence the Yeoh model was chosen for the separate fascia and the determined parameter values were used (Table 1). In order to find the parameters for the combined fascia-muscle material, the calibration load case was used (Figure 3), which is similar to a flat surface compression test. It was also observed during these calibration tests, that using muscle tissue properties for the fascia, as if the fascia was ignored, yielded a softer homogenized response than if the fascia was given the combined properties (used for the Combined material set) or given material data for fascia (Separate set).

This indicated that ignoring fascia properties compared to using the Separate material set would result in larger differences compared to if the fascia was combined with muscle properties, as in the Combined set. Based on these findings under these material conditions, it is of importance to consider fasciae properties and separate the types of soft tissues in the lower limb cross section models for FEA simulations. However, in order to evaluate a socket design choice by simulation requires more information regarding the individual specific case, biomechanical loading conditions, individual specific geometries and material properties, as well as individual tolerance to loads in specific tissues etc. Further evaluations of different material models for specific tissues would also be of interest.

Potential problems in our geometric model of soft tissues were that some locations of the tissue boundaries had sharper corners than the real anatomy and some elements in the finite element mesh had small corner angles. Large elements and sharp corner angles increase the stiffness of those elements. The latter issue was accounted for when evaluating results by excluding single nodal extreme values. The mesh where the fascia region was very thin, as at the bones, was refined to small elements. A smoother geometry to avoid sharp corners on curvatures and refined mesh in areas of concern would improve the models further. For the purpose of this study, the lower limb model geometry was the same for all simulations, and thus comparisons of material set, and external socket shapes were deemed useful.

The tibia was fixed while the fibula could move based on reported experimental findings [44, 45], and we did not assume an amputation with a bone bridge. Whether a transtibial bone bridge technique improves residual limb loading ability is yet unclear [49] thus simulation studies would be of interest. Currently, we have only investigated the internal mechanical conditions due to the shape constraints of sockets without considering the loading due to body weight or dynamics. Contact pressure was computed as a result of the model configuration. This strategy avoided constraints that might have arisen when contact forces are applied as boundary conditions. In an amputated leg, due to the weight of the subject and changing cross section along the length of the leg, the stress-strain distribution is quite complex. This simulation for conditions of soft tissues at rest without load bearing or active muscle contractions showed that there are still meaningful differences in outcomes at these considered low loads. Maxima sites and magnitudes per five soft tissue types and contact pressures between skin and sockets varied by material set and socket design used. The relative change between the use of material sets varied clearly. No such comparison of tissue representation were found in literature elsewhere. Comparisons between studies are limited due to the differences in simulation models [13, 18, 21], and thus a standard model such as the one used here, would be of interest. We believe that more realistic representation of separate soft tissues is needed in simulation models with FEA, and that careful interpretations of simulation results are necessary.

**Achievements**

This paper presents

- a detailed geometrical human transtibial model with 6 specified tissues; skin, adipose fat, fasciae, muscles, blood vessels and bones,
- effects of considering fascia as a separate tissue or as combined with muscle tissues,
- the level of segmental tissue type representation mattered; the magnitudes and sites of maxima in each tissue type varied with material set,
- computed contact forces at skin-socket interface differed by the material set and the socket design used, and,

- the relative changes in outcomes were considerable with the Combined material set of 5 tissue types in relation to the Separate set with 6 types.

**Conclusion**

This study showed that the two representations of fascia and muscle tissues influenced the results with considerable differences. Thus, the level of detail for tissues in lower leg models need to be considered when creating models and interpreting results of FEA simulations. Further developments and evaluations of simulation models of the human lower limb tissues are needed.

**Conflict of interest**

The authors declare no conflict of interest. The founding sponsors had no role in the design of the study; in the collection, analyses, or interpretation of data; in the writing of the manuscript, and in the decision to publish the results.

**Acknowledgements**

Hotswap Norden AB, Ottobock Scandinavia AB and Ossur Nordic HF are greatly acknowledged for their partnership in the PEOPLE project: PrEvention Of Pressure uLcers and dEep tissue injury by optimization of body tight external supports.

Financial support from the Jönköping Regional Research Programme, Sweden, to the project PEOPLE is greatly acknowledged.

**Author contributions**

S Kallin, A Rashid and K Salomonsson conceived and designed the study; S Kallin created the socket geometries; A Rashid created the geometric models and contact problem; S Kallin performed the numerical experiments and analysed the data supervised by K Salomonsson and P Hansbo; S Kallin was the main author of the paper.